\documentclass[twocolumn,pre]{revtex4}

\usepackage{dcolumn}
\usepackage{amsmath}

\usepackage{graphicx}
\usepackage{bm}
\usepackage{calrsfs}

\begin{document}

% Macros
\newcommand{\e}{\mathrm{e}}
\renewcommand{\d}{\mathrm{d}}
\newcommand{\Ord}{\mathrm{O}}
\newcommand{\eref}[1]{(\ref{#1})}
\newcommand{\etal}{\textit{et~al.}}
\newcommand{\half}{\mbox{$\frac12$}}
\newcommand{\av}[1]{\left\langle#1\right\rangle}
\newcommand{\set}[1]{\left\lbrace#1\right\rbrace}
\newcommand{\cH}{\mathcal{H}}
\newcommand{\cD}{\mathcal{D}}
\newcommand{\bM}{\mathbf{M}}
\newcommand{\bQ}{\mathbf{Q}}
\newcommand{\vphi}{\bm{\phi}}
\newcommand{\vxi}{\bm{\xi}}

% Style parameters
\setlength{\parskip}{0pt}
\setlength{\tabcolsep}{6pt}
\setlength{\arraycolsep}{2pt}

\title{Solution of the 2-star model of a network}
\author{Juyong Park}
\affiliation{Department of Physics, University of Michigan,
Ann Arbor, MI 48109--1120}
\author{M. E. J. Newman}
\affiliation{Department of Physics, University of Michigan,
Ann Arbor, MI 48109--1120}
\begin{abstract}
The $p$-star model or exponential random graph is among the oldest and
best-known of network models.  Here we give an analytic solution for the
particular case of the 2-star model, which is one of the most fundamental
of exponential random graphs.  We derive expressions for a number of
quantities of interest in the model and show that the degenerate region of
the parameter space observed in computer simulations is a spontaneously
symmetry broken phase separated from the normal phase of the model by a
conventional continuous phase transition.
\end{abstract}
\maketitle

\section{Introduction}
There has in recent years been a surge of interest within the physics
community in the properties of networks, including the Internet, the world
wide web, and social and biological networks of various
kinds~\cite{Strogatz01,AB02,Newman03d,DM03b}.  Work has been divided
between studies of specific real-world networks, along with the development
of measures and algorithms for their analysis, and the creation of models
to predict and explain network behavior.  It is on models that we focus
here.

Network modeling goes back at least as far as the well-known random graph
or Bernoulli graph, studied by Solomonoff and Rapoport in the early
1950s~\cite{SR51} and famously by Erd\H{o}s and R\'enyi~\cite{ER60} a
decade later.  The random graph however is a poor model for most real-world
networks, as has been argued by many authors~\cite{Strogatz01,DM03b,WS98},
and so other models have been developed.  Recent attention has focused
particularly on generalized random graphs such as the configuration
model~\cite{MR95,ACL00,NSW01} and on generative models, particularly models
of growing networks~\cite{AB02,DM03b,BA99b,KR03}.  There is, however,
another class of network models that, while widely used and valuable, has
attracted little attention in the physics community, namely the class of
``exponential random graphs'' or ``$p$-star models.''  Building on early
statistical work by Besag~\cite{Besag74}, exponential random graphs were
first studied in the 1980s by Holland and Leinhardt~\cite{HL81}, and later
developed extensively by Strauss and others~\cite{Strauss86,FS86}.  Today,
they are commonly used as a practical tool by statisticians and social
network analysts~\cite{WP96,AWC99,Snijders02}.

Despite their widespread adoption, few analytic results are known for
exponential random graphs: most work has made use of computer simulation to
fit models to observational data and evaluate model predictions.
Exponential random graphs however are ideally suited to study using the
techniques of statistical physics.  Recently, physicists have examined
exponential random graph models of network assortativity~\cite{BL02,PN03}
and transitivity~\cite{BJK04a}.  Here we take a different approach and show
how physics techniques can be used to derive analytically the behavior of
one of the most fundamental of exponential random graph models, the 2-star
model.  We view this solution not only as a calculation of interest in its
own right, but also as a demonstration of the way in which physics
techniques can be fruitfully applied to problems from other fields.

\section{The model}
The exponential random graph is an ensemble model.  One defines an ensemble
consisting of the set of all simple undirected graphs with $n$ vertices and
no self-edges (i.e.,~networks with either zero or one edge between each
pair of distinct vertices) and one specifies a probability $P(G)$ for each
graph~$G$ in this ensemble.  Properties of the model are calculated as
averages over the ensemble.  Let us define the \textit{graph Hamiltonian},
also referred to by statisticians as a \textit{log odds ratio}, to be $H(G)
= F - \ln P(G)$.  Here $F$ (usually called the free energy) is any
convenient origin for the measurement of the Hamiltonian, such as, for
instance, the log of the probability of the empty graph (i.e.,~the
probability of $n$~vertices with no edges).  Then
\begin{equation}
P(G) = {\e^{-H(G)}\over Z},\qquad Z = \e^{-F} = \sum_G \e^{-H(G)}.
\label{boltzmann}
\end{equation}
$Z$ is the graph partition function and many quantities of interest can be
calculated from it, or alternatively from the free energy.

So far, this model is entirely general, but progress is made by assuming
the Hamiltonian to be a linear combination of scalar graph observables,
such as number of edges, degree sequences, or clustering coefficients.  In
this paper we study one of the simplest nontrivial cases, the 2-star model,
for which $H(G) = \theta_1 m(G) + \theta_2 s(G)$, where $\theta_1$ and
$\theta_2$ are independent parameters, $m(G)$ is the number of edges in the
graph and $s(G)$ is the number of ``2-stars.''  A 2-star is a pair of edges
that share a common vertex.

Let us denote by $k_i$ the degree of vertex~$i$.  Then $m(G)=\frac12\sum_i
k_i$ and $s(G)=\half\sum_i k_i(k_i-1)$, and hence we can write the
Hamiltonian in the form
\begin{equation}
H = -{J\over n-1}\sum_i k_i^2 - B\sum_i k_i,
\label{defsh}
\end{equation}
where the ``coupling constant'' $J=-\half(n-1)\theta_2$ and the ``field''
$B=\half(\theta_2-\theta_1)$.  The factor $(n-1)$ in the definition of $J$
is not strictly necessary, but it makes the equations simpler later on.

There are two analytic approaches from statistical mechanics that can be
brought to bear on problems like this.  The first is to use perturbation
theory~\cite{BJK04a} and the second is to use non-perturbative techniques,
usually based on the Hubbard--Stratonovich transform and saddle-point
expansions~\cite{BL02}.  Here we make use of the latter to solve the 2-star
model.

\section{Analytic approach}
Our goal is to calculate the partition function~$Z$, Eq.~\eref{boltzmann},
or equivalently the free energy.  First, we introduce auxiliary
fields~$\phi_i$ on the vertices of the graph using the
Hubbard--Stratonovich relation
\begin{eqnarray}
\exp\bigl(Jk_i^2/(n-1)\bigr) &=& \sqrt{(n-1)J\over\pi} \nonumber\\
   & & \hspace{-8em} \times \int_{-\infty}^\infty\d\phi_i\,
       \exp\bigl(-(n-1)J\phi_i^2+2J\phi_i k_i\bigr),
\end{eqnarray}
which gives
\begin{eqnarray}
Z &=& \biggl[{(n-1)J\over\pi}\biggr]^{n/2}
    \int\cD\vphi\>\exp\Bigl(-(n-1)J\sum_i\phi_i^2\Bigr)\nonumber\\
  & & \quad\times\sum_G \exp\Bigl(\sum_i (2J\phi_i+B)k_i\Bigr),
\label{partition1}
\end{eqnarray}
where $\cD\vphi$ indicates the path integral over the fields~$\set{\phi_i}$
and we have interchanged the order of the integral and the sum over
graphs~$G$.

The sum over graphs can now be performed by defining the symmetric
adjacency matrix $\sigma_{ij}$ equal to 1 if there is an edge between
vertices $i$ and $j$ and zero otherwise.  Then, noting that $k_i = \sum_j
\sigma_{ij}$, we can write
\begin{eqnarray}
\sum_i (2J\phi_i+B)k_i
  &=& \sum_{ij} (2J\phi_i+B)\sigma_{ij}\nonumber\\
  &=& \sum_{i<j} \bigl[ 2J(\phi_i+\phi_j) + 2B \bigr] \sigma_{ij}.
\end{eqnarray}
Since $\sigma_{ij}$ is symmetric, its values for $i<j$ completely define
the graph, and hence
\begin{eqnarray}
\sum_G \exp\Bigl(\sum_i (2J\phi_i+B)k_i\Bigr)
  &=& \prod_{i<j} \sum_{\sigma_{ij}=0}^1
      \e^{[2J(\phi_i+\phi_j)+2B]\sigma_{ij}}\nonumber\\
  & & \hspace{-6em} =
      \prod_{i<j} \bigl(1 + \e^{2J(\phi_i+\phi_j)+2B}\bigr).
\end{eqnarray}
Substituting this result into Eq.~\eref{partition1}, we then get
\begin{equation}
Z = \int\cD\vphi\>\e^{-\cH(\vphi)},
\label{ft}
\end{equation}
where the effective Hamiltonian~$\cH$ is
\begin{eqnarray}
\cH(\vphi) &=& (n-1)J\sum_i \phi_i^2 - \half\sum_{i\ne j}
                  \ln \bigl( 1 + \e^{2J(\phi_i+\phi_j)+2B}\bigr) \nonumber\\
           & & {} - \half n\ln\bigl((n-1)J\bigr).
\label{effective}
\end{eqnarray}

Thus we have transformed our network model into a field theory of a
continuous scalar field on $n$ sites, which can be solved using a variety
of methods.  The simplest mean-field approach is to ignore fluctuations and
assume $\phi_i$ always to be equal to its most probable value, which occurs
at the saddle point
\begin{equation}
{\partial\cH\over\partial\phi_i} = 0
  = 2(n-1)J\phi_i - J\sum_{j(\ne i)}
    \bigl[ \tanh\bigl(J(\phi_i+\phi_j)+B\bigr) + 1 \bigr].
\label{mft1}
\end{equation}
This has a symmetric solution $\phi_i=\phi_0$ for all~$i$ with
\begin{equation}
\phi_0 = \half\bigl[ \tanh\bigl(2J\phi_0+B\bigr) + 1 \bigr].
\label{mft2}
\end{equation}

\begin{figure}
\begin{center}
\resizebox{8.5cm}{!}{\includegraphics{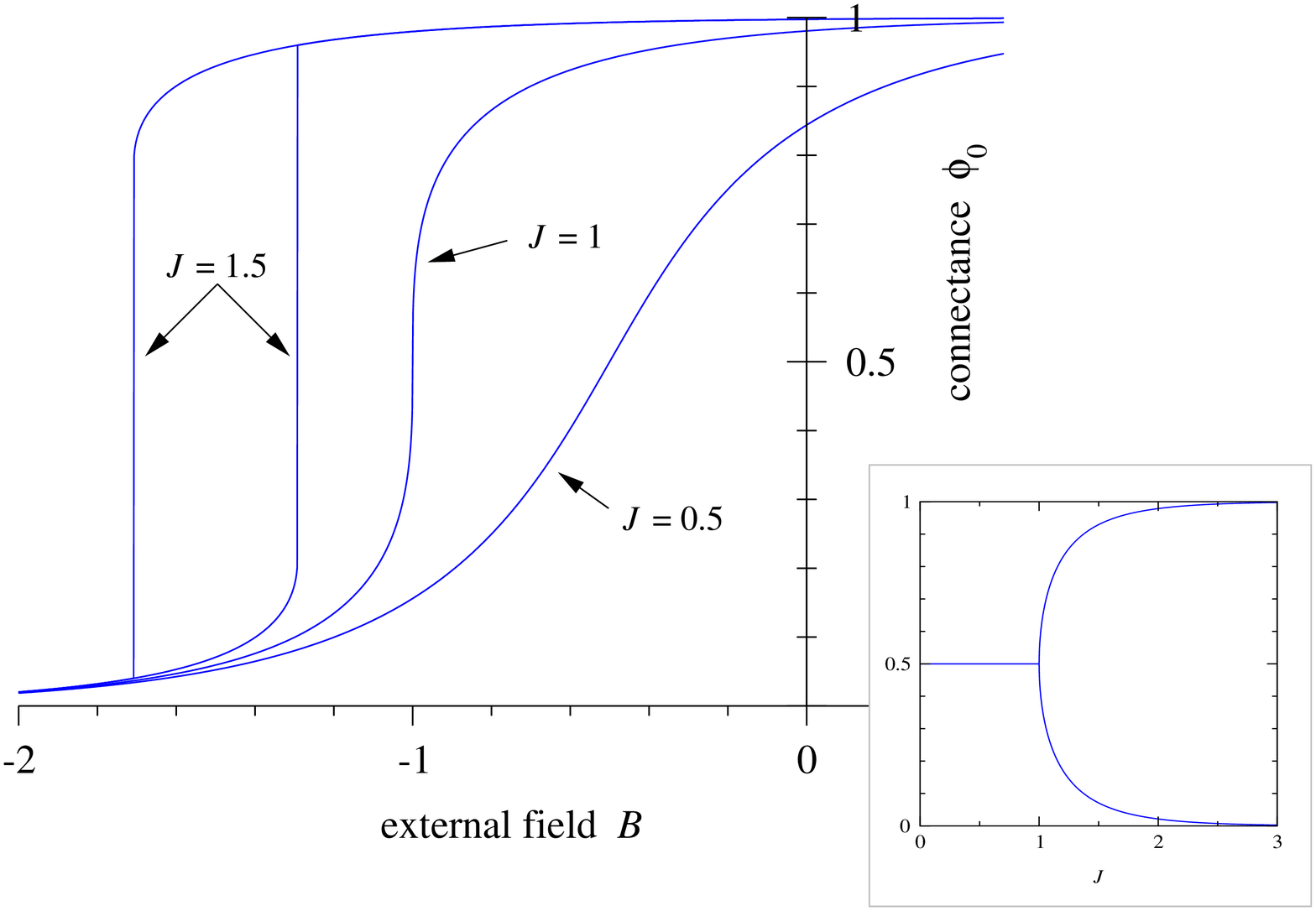}}
\end{center}
\caption{The mean-field solution for the connectance $\phi_0=\av{k}/(n-1)$
in the 2-star model from Eq.~\eref{mft2}, for values of the coupling $J$
below, at, and above the phase transition.  For the case $J=1.5$ we are in
the symmetry broken phase and the hysteresis loop corresponding to the
high- and low-density phases of the system is clearly visible.  Inset: the
bifurcation of the connectance as a function of~$J$ along the symmetric
line $B=-J$.}
\label{meanfield}
\end{figure}

This quantity has a simple physical interpretation.  The mean
degree~$\av{k}$ of a vertex in the graph is given by the derivative of the
free energy thus:
\begin{eqnarray}
\av{k} &=& {1\over n}\sum_i\av{k_i}
        =  {1\over n} {\partial F\over\partial B} \nonumber\\
       &=& {1\over2n} \sum_{i\ne j}
           \av{\tanh\bigl(J(\phi_i+\phi_j)+B\bigr) + 1}_\phi,
\end{eqnarray}
where $\av{\dots}_\phi$ indicates an average in the $\vphi$ ensemble of
Eq.~\eref{ft}.  Making the mean-field assumption of Eq.~\eref{mft2}, this
becomes $\av{k}=(n-1)\phi_0$ and hence $\phi_0$ is simply proportional to
the mean degree of a vertex, within the mean-field approximation.  The
quantity $\av{k}/(n-1)$ is called the ``connectance'' of the graph---it is
the fraction of possible edges that are actually present and is a measure
of the mean density.  So we could also say that $\phi_0$ is equal to the
connectance.  This allows us to interpret Eq.~\eref{mft2} very directly.
For $J\le1$, this equation has only a single solution, but for $J>1$ we
have three coexisting solutions when $B$ is sufficiently close to~$-J$.
Only the outer two solutions are stable, giving us a bifurcation at $J_c=1$
corresponding to a continuous phase transition at this point to a symmetry
broken state exhibiting two phases, one of high density (typically nearly a
complete graph) and one of low density.  We show a plot of the solution
of~\eref{mft2} in the main panel of Fig.~\ref{meanfield}.

Along the line $B=-J$ the Hamiltonian~\eref{defsh} is symmetric with
respect to the interchange of edges and ``holes''---the absence of edges
between vertex pairs.  In the inset to Fig.~\ref{meanfield} we show the
solution for the connectance as a function of $J$ along this symmetric
line and the plot shows the bifurcation clearly.

To move beyond the mean-field result, we make use of the method of
stationary phase.  Expanding the effective Hamiltonian~\eref{effective}
about the mean-field solution to leading order we have
$\cH=\cH(\vphi_0)+\vphi'\bM\vphi'+\Ord(\vphi^3)$, where
$\vphi'\equiv\vphi-\vphi_0$ and $\bM$ is the Hessian matrix of second
derivatives of $\cH$ with respect to~$\vphi$, evaluated at~$\vphi_0$.
Changing variables to $\vxi=\bQ\vphi'$, where $\bQ$ is the matrix of
eigenvectors of~$\bM$, $\bM$~is diagonalized and
$\cH=\cH(\vphi_0)+\sum_i\lambda_i\xi_i^2+\Ord(\vxi^3)$, with $\lambda_i$
being the $i$th eigenvalue of~$\bM$.  Substituting into Eq.~\eref{ft} and
observing that the Jacobian of the variable change $|\bQ|=1$, the path
integral becomes a product of independent Gaussian integrals and $Z =
\e^{-\cH(\vphi_0)}/\sqrt{|\bM|}$, or equivalently $F = \cH(\vphi_0) +
\half\ln|\bM|$, where $|\bM|$ is the determinant of~$\bM$.

The elements of the Hessian matrix have the values:
\begin{equation}
M_{ij} = \biggl\lbrace\begin{array}{ll}
           -4J^2 \phi_0(1-\phi_0)
             & \quad\mbox{for $i\ne j$,} \\
	   (n-1) [2J - 4J^2 \phi_0(1-\phi_0)]
             & \quad\mbox{for $i=j$,}
         \end{array}
\end{equation}
giving
\begin{equation}
|\bM| = (2(n-1)J)^n \bigl( 1-2J \phi_0(1-\phi_0) \bigr)^{n-1}
                    \bigl( 1-4J \phi_0(1-\phi_0) \bigr).
\label{determinant}
\end{equation}
Then, making use of Eqs.~\eref{effective} and~\eref{mft2}, we arrive at the
solution for the free energy
\begin{eqnarray}
F &=& n(n-1)J\phi_0^2 - \half n(n-1) \ln\bigl( 1 + \e^{4J\phi_0+2B} \bigr)
        \nonumber\\
  & & \quad {} + \half(n-1)\ln\bigl(1-2J\phi_0(1-\phi_0)\bigr),
\end{eqnarray}
where we have kept leading order corrections to the mean-field result but
dropped terms of order a constant and smaller that vanish in the large~$n$
limit.

\begin{figure}
\begin{center}
\resizebox{7.5cm}{!}{\includegraphics{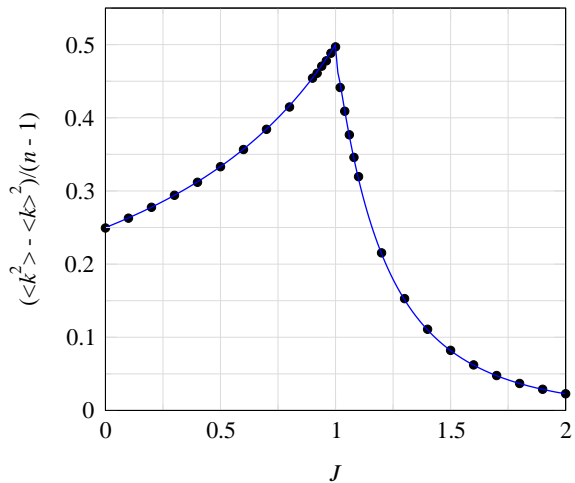}}
\end{center}
\caption{The variance of vertex degree in the 2-star model as a function of
the coupling~$J$ along the symmetric line $B=-J$.  The phase transition is
marked by a cusp in the variance, but no divergence.  The solid line
represents the analytic solution, Eq.~\eref{variance}, in the large system
size limit, and the points are the results of Monte Carlo simulations of
the model for $n=1000$.}
\label{cusp}
\end{figure}

From the free energy we can calculate expected values of a variety of
properties of the model.  For instance the mean degree~$\av{k}$ and the
mean squared degree~$\langle k^2 \rangle$ are given by derivatives with
respect to $B$ and $J$ and are equal to
\begin{eqnarray}
\label{spk}
\av{k} &=& (n-1)\phi_0 + \nonumber\\
       & & \hspace{-1em}\frac{2J\phi_0(1-\phi_0)(1-2\phi_0)}{\bigl(1-4J\phi_0(1-\phi_0)\bigr)\bigl(1-2J\phi_0(1-\phi_0)\bigr)},
         \\
\label{spksq}
\av{k^2} &=& (n-1)^2 \phi_0^2 + \nonumber\\
         & & \hspace{-1em}\frac{(n-1)\phi_0(1-\phi_0)(1-4J\phi_0^2)}{(1-4J\phi_0(1-\phi_0))(1-2J\phi_0(1-\phi_0))}.
\end{eqnarray}
The leading order term in each case is the same as the mean-field result,
so that in the limit of large~$n$ both $\av{k}$ and $\langle k^2 \rangle$
take their mean-field values.  The variance of the degree $\langle k^2
\rangle-\av{k}^2$ on the other hand is zero within the mean-field
approximation because of the cancellation of the leading terms but non-zero
beyond mean-field:
\begin{equation}
\langle k^2 \rangle - \av{k}^2 =
  (n-1)\frac{\phi_0(1-\phi_0)}{1-2J\phi_0(1-\phi_0)}.
\label{variance}
\end{equation}
From consideration of Fig.~\ref{meanfield} one might expect this quantity
to diverge at the phase transition, but in fact it does not, having merely
a cusp at that point.  In Fig.~\ref{cusp} we show the form of this function
along the symmetric line $B=-J$ as a function of~$J$.  The figure also
shows the results of Monte Carlo simulations of the 2-star model for the
same parameter values and, as we can see, agreement between the simulations
and the analytic solution is excellent.

A divergence does occur in the variance of the number of edges in the
network at the phase transition.  This quantity, which plays the role of a
susceptibility for the model, is given to leading order by
\begin{equation}
\langle m^2 \rangle - \av{m} = {\partial^2 F\over\partial B^2}
   = (n-1) {2\phi_0(1-\phi_0)\over 1 - 4J\phi_0(1-\phi_0)}.
\end{equation}
This diverges as $|J-J_c|^{-1}$ as we approach the transition along the
symmetric line $B=-J$.

One can also ask whether the network described by the 2-star model
possesses a giant component.  Molloy and Reed~\cite{MR95} have demonstrated
that a network without degree correlations possesses a giant component if
and only if $\langle k^2 \rangle>2\av{k}$.  We can evaluate this criterion
using Eqs.~\eref{spk} and~\eref{spksq}, and find that for all values of the
system parameters the network possesses a giant component in the limit of
large~$n$.

\begin{figure}
\begin{center}
\resizebox{7.5cm}{!}{\includegraphics{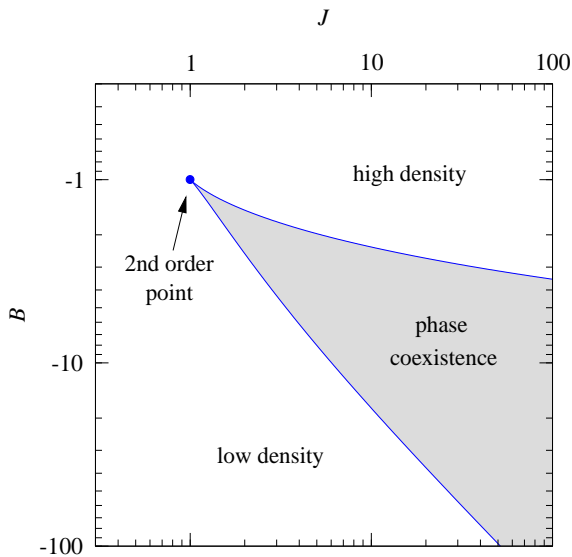}}
\end{center}
\caption{The phase diagram for the 2-star model.  The shaded region
indicates the hysteretic region in which both high- and low-density phases
are possible.}
\label{phasediagram}
\end{figure}

In Fig.~\ref{phasediagram} we show the phase diagram for the 2-star model
as a function of the parameters $J$ and~$B$.  The critical point is at
$J=1$, $B=-1$ and beyond this point there are high- and low-density phases
separated by a phase coexistence region.  In the coexistence region the
phase of the model depends on its history in a manner characteristic of
hysteretic systems.  Some studies of exponential random graphs have
considered the case in which the number of edges in the graph is fixed, a
``conserved-order-parameter'' version of the current model~\cite{BL02}.  In
such a case, the phase coexistence region will correspond to true
coexistence; low free-energy states of the system will be states in which
the system prefers simultaneously to have some high-degree ``hub'' vertices
that connect to essentially all others and some of lower degree, rather
than being uniform everywhere.  Such ``degenerate'' behavior has been
observed since the earliest numerical experiments on exponential random
graphs~\cite{HL81,Strauss86,FS86,PDFV04}.  Here we see that this behavior
is the precise network analog of the phase separation phenomenon known to
physicists from many other systems.

\section{Conclusions}
In this paper, we have given a non-perturbative analytic solution of one of
the oldest of network models, the 2-star model, which is perhaps the
simplest nontrivial model of the class known as exponential random graphs
and has been long studied in the social sciences.  The model turns out to
be perfectly suited to solution by the methods of statistical physics, and
among other things the solution shows the degenerate behavior of the model
in certain parameter regimes to be the result of a symmetry breaking
between high- and low-density phases, which are separated from the
``normal'' region of the model by a continuous phase transition.

The exponential random graphs are, we believe, an important class of
network models, which have largely been neglected despite the high level of
interest in networks in the last few years.  We hope that others will also
take up the study of these models, either using methods like those
discussed here or other methods yet to be described.

\begin{acknowledgments}
The authors thank Julian Besag, Mark Handcock, and Pip Pattison for useful
conversations.  This work was supported in part by the National Science
Foundation under grant number DMS--0234188 and by the James S. McDonnell
Foundation.
\end{acknowledgments}

%\bibliographystyle{numeric}
%\bibliography{journals,references}

\end{document}